\documentclass{PoS}
\usepackage{graphicx}
\usepackage{amsmath}

\RequirePackage{xspace}
\def\epem {\ensuremath{e^+e^-}\xspace}
\newcommand{\gev}{\ensuremath{\mathrm{\,Ge\kern -0.1em V}}\xspace}
\newcommand{\gevcc}{\ensuremath{{\mathrm{\,Ge\kern -0.1em V\!/}c^2}}\xspace}
\newcommand{\tev}{\ensuremath{\mathrm{\,Te\kern -0.1em V}}\xspace}
\def\km   {\ensuremath{{\rm \,km}}\xspace}

\def\mm   {\ensuremath{{\rm \,mm}}\xspace}
\def\mum  {\ensuremath{{\,\mu\rm m}}\xspace}
\def\nm   {\ensuremath{{\rm \,nm}}\xspace}
\def\c  {\ensuremath{{\rm \,c}}\xspace}
\def\cms  {\ensuremath{{\rm \,cm}^{-2} {\rm s}^{-1}}\xspace}
\newcommand{\lum} {\ensuremath{\mathcal{L}}\xspace}

\title{\boldmath Limitation on the luminosity of \epem
  storage rings due to beamstrahlung}

\ShortTitle{Limitation on the luminosity of \epem
  storage rings due to beamstrahlung}

\author{\speaker{V.~I.~Telnov}\\
        Budker Institute of Nuclear Physics, SB RAS, 630090, Novosibirsk, Russia\\
        Novosibirsk State University, 630090, Novosibirsk, Russia\\
        E-mail: \email{telnov@inp.nsk.su}}

\abstract{Particle loss due to the emission of single energetic beamstrahlung photons in beam collisions is shown to impose a fundamental limit on storage-ring luminosities at energies greater than $2E_0 \sim 140 \gev$ for head-on collisions and $2E_0 \sim 40 \gev$ for crab-waist collisions. Above these threshold energies, the suppression factor due to beamstrahlung scales as $1/E_0^{4/3}$, and for a fixed power of synchrotron radiation, the luminosity $\lum \propto R/E_0^{13/3}$, where $R$ is the collider radius. For $2E_0 \gtrsim 150 \gev$, both collision schemes have similar luminosity limits. The luminosities attainable at storage-ring and linear-collider (LC) $2E_0 \sim 240 \gev$ Higgs factories are comparable; at higher energies, LCs are preferable.
This conference paper is based on my recent PRL publication~\cite{Telnov-prl} supplemented with additional comments on linac-ring \epem colliders and ring \epem colliders with charge compensation (four-beam collisions).
}

\FullConference{LHC on the March --- IHEP--LHC,\\
        20--22 November 2012\\
        Institute for High Energy Physics, Protvino, Moscow region, Russia }
\PACS{29.20}

\begin{document}

\section{Introduction} \noindent
The ATLAS and CMS experiments at the LHC recently discovered~\cite{higgsATLAS,higgsCMS}
the long-sought Higgs boson with the mass $M \approx 126\gevcc$.
The precision study of the Higgs boson's properties would require the construction of an energy- and luminosity-frontier \epem collider~\cite{Aarons}.

With the sole exception of the $2E_0=91 \gev$ SLAC Linear Collider (SLC),
all \epem colliders ever built have been based on storage rings.
The $2E_0=209 \gev$ LEP collider at CERN is generally considered to have been the last energy-frontier \epem storage-ring collider (SRC).
Indeed, synchrotron-radiation energy losses, which are proportional to
$E_0^4/R$, make the construction and operation of higher-energy SRCs excessively expensive.
Energy-frontier linear \epem colliders (LC), which have been in development for over 40 years,
are free from this limitation and allow multi-\tev energies to be reached.
Two LC projects are in advanced stages of development: the $2E_0=500 \gev$ ILC~\cite{ILC} and
the $2E_0=500$--3000\gev CLIC~\cite{CLIC}.

Nevertheless, several proposals~\cite{Zim,Oide,Blo,Sum,Blo2} for a $2E_0=240 \gev$ SRC
for the study of the Higgs boson in $\epem \to HZ$ have recently been put forward~\cite{Sen}.
Lower cost and reliance on firmly established technologies are cited as these projects' advantages over an LC. Moreover, it has been proposed~\cite{Oide} that a $2E_0=240 \gev$ SRC can provide
superior luminosity, and that the ``crab waist'' collision scheme~\cite{crab,SuperB} can be adapted
to energy-frontier SRCs, allowing them to exceed the ILC and CLIC luminosities
even at $2E_0=400$--500 \gev. Parameters of  SRCs proposed in Refs.~\cite{Zim,Oide} are summarized in Table~\ref{Table1}.

In this conference paper, which is based on \cite{Telnov-prl}, we examine the role of {\sl beamstrahlung},
i.e., synchrotron radiation in the electromagnetic field of the opposing beam, in high-energy \epem SRCs.
First discussed in~\cite{Rees}, beamstrahlung has been well-studied only in the LC case~\cite{Chen}.
As we shall see, at energy-frontier \epem SRCs beamstrahlung determines the beam lifetime through the emission of single photons in the tail of the beamstrahlung spectra, thus severely limiting the
luminosity. Unlike the LC case, beamstrahlung has little effect on the SRC energy spread.
\begin{table*}[!hbt]
\small
{\renewcommand{\arraystretch}{0.75}
\begin{tabular}{l c c c c c c c c c c c}  \\[-2.5mm]
& LEP & LEP3 & DLEP & STR1 & STR2 & STR3 &  STR4 &  STR5 &  STR6 \\[-1mm]
&&&&&&cr-w&cr-w&cr-w&cr-w \\
\hline \\[-2.2mm]
$2E_0$, \gev      & 209 & 240 & 240 & 240 & 240 & 240 & 400 & 400 & 500   \\
Circumference,\km & 27& 27 & 53 & 40 & 60 & 40 & 40 & 60 & 80 \\
Beam current, mA  & 4 & 7.2 & 14.4 & 14.5 & 23 & 14.7 & 1.5 & 2.7 & 1.55 \\
Bunches/beam & 4  & 3 & 60 & 20 & 49 & 15 & 1 & 1.4 & 2.2 \\
$N,\;10^{11}$          & 5.8 & 13.5 & 2.6 & 6 & 6 & 8.3 & 12.5 & 25.& 11.7  \\
$\sigma_{z}$,\mm    & 16  & 3   & 1.5 & 3   &  3 & 1.9 & 1.3 & 1.4 & 1.9  \\
$\varepsilon_x$,\nm   & 48   & 20   & 5  & 23.3& 24.6& 3   & 2  & 3.2& 3.4\\
$\varepsilon_y$,\nm   & 0.25 & 0.15 & 0.05 & 0.09 & 0.09 & 0.011 & 0.011 & 0.017 & 0.013\\
$\beta_x$,\mm      & 1500 & 150 & 200 & 80  & 80  & 26  & 20 & 30 & 34 \\
$\beta_y$,\mm      & 50 & 1.2 & 2 & 2.5 & 2.5 & 0.25 & 0.2 & 0.32 & 0.26 \\
$\sigma_x$,\mum & 270 & 54   & 32 & 43  & 44  & 8.8 & 6.3 & 9.8 & 10.7  \\
$\sigma_y$,\mum & 3.5 & 0.42 & 0.32 & 0.47 & 0.47 & 0.05 & 0.047 & 0.074 & 0.06 \\
SR power, MW       & 22  & 100  & 100 & 100  & 100  & 100 & 100 & 100 & 100 \\
Energy loss per turn,\gev & 3.4 & 7 & 3.47 & 3.42 & 2.15 & 3.42 & 33.9 & 18.5 & 32.45 \\
\lum, $10^{34}\cms$ & 0.013 & 1.3 & 1.6 & 1.7 & 2.7 & 17.6 & 4 & 7 & 2.2 \\
\hline \hline \\[-2.6mm]
$E_\mathrm{c,max}/E_0$,$\, 10^{-3}$ & 0.09 & 6.3 & 4.2 & 3.5 & 3.4 & 38 & 194 & 232 & 91 \\
$n_{\gamma}$ per electron    & 0.09 & 1.1 & 0.37 & 0.61 & 0.6 & 4.2 & 8.7&  11.3 & 4.8 \\
lifetime(SR@IP), s & $\sim \infty$ & 0.02 & 0.3 & 0.2 & 0.4 & 0.005 & 0.001 & 0.0005 & 0.005 \\
\hline \\[-2.6mm]
$\lum_\mathrm{corr}$, $10^{34}\cms$ & 0.013 & 0.2 & 0.4 & 0.5 & 0.8 & 0.46 & 0.02 & 0.03 & 0.024 \\
\end{tabular}
\vspace{-0mm}
}
\caption{Parameters of LEP and several recently proposed storage-ring
colliders~\cite{Zim,Oide}.
``STR'' refers to ``SuperTRISTAN''~\cite{Oide}.
Use of the crab-waist collision
scheme~\cite{crab,SuperB} is denoted by ``cr-w''.
 The luminosities and the numbers of bunches for all projects
 are normalized to the total synchrotron-radiation power of 100 MW.
Beamstrahlung-related quantities
derived in this paper are listed below the double horizontal line.}
\label{Table1}
\end{table*}

At LCs, flat beams are employed to suppress beamstrahlung.
Each colliding particle radiates $n_{\gamma}=1$--2 beamstrahlung photons with the total
energy averaging 3--5\% of the beam energy.
The long tails of the beamstrahlung energy-loss spectrum are not a
problem for LCs because beams are used only once.

In contrast, at SRCs the particles that lose a certain fraction
of their energy in a beam collision leave the beam and strike
the vacuum chamber's walls; this fraction $\eta$ is typically around 0.01
(0.012 at LEP) and is known as the ring's energy acceptance.
Beamstrahlung was negligible at all previously built SRCs because of
their relatively large beam sizes. Its importance considerably increases
with energy. Table~\ref{Table1} lists the beamstrahlung characteristics
of the newly proposed SRCs assuming a 1\% energy acceptance:
the critical photon energy for the maximum
beam field $E_\mathrm{c,max}$, the average number of beamstrahlung photons per
electron per beam crossing $n_{\gamma}$, and the beamstrahlung-driven
beam lifetime.
Please note that once beamstrahlung is taken into
account, the beam lifetimes drop to unacceptably low values, from a
fraction of a second to as low as a few revolution periods.

At the SRCs considered in Table~\ref{Table1}, the beam lifetime due to
the unavoidable radiative Bhabha scattering is 10 minutes or longer.
One would therefore want the beam lifetime due to beamstrahlung to be
at least 30 minutes. The simplest (but not optimum) way to suppress
beamstrahlung is to decrease the number of particles per bunch with a
simultaneous increase in the number of colliding bunches.  As
explained below, $E_\mathrm{c,max}$ should be reduced to $\approx
0.001E_0$. Thus, beamstrahlung causes a great drop in luminosity,
especially at crab-waist SRCs: compare the proposed \lum and corrected
(as suggested above) $\lum_\mathrm{corr}$ rows in Table~\ref{Table1}.

To achieve a reasonable beam lifetime, one must make small
the number of beamstrahlung photons with energies greater than
the threshold energy $E_\mathrm{th}=\eta E_0$ that causes the electron
to leave the beam. These photons belong to the high-energy
tail of the beamstrahlung spectrum and have energies much greater
than the critical energy. It will be clearly shown below that the
beam lifetime is determined by such single high-energy
beamstrahlung photons, not by the energy spread due to the emission
of multiple low-energy photons.

\section{Beam lifetime due to beamstrahlung, restriction on beam parameters~{\rm \cite{Telnov-prl}}}
\noindent
The critical energy for synchrotron radiation
\begin{equation}
E_\mathrm{c}=\hbar \omega_\mathrm{c} = \hbar\frac{3\gamma^3c}{2\rho},
\label{1}
\end{equation}
where $\rho$ is the bending radius and $\gamma=E_0/mc^2$. The
spectrum of photons per unit length with energy well above the
critical energy
\begin{equation}
\frac{dn}{dx} = \sqrt{\frac{3\pi}{2}}\frac{\alpha \gamma}{2\pi \rho}
\frac{e^{-u}}{\sqrt{u}}du,
\label{2}
\end{equation}
where $u=E_{\gamma}/E_\mathrm{c}$, $\alpha=e^2/\hbar c$.

Using these simple formulas, one can find the critical energy of beamstrahlung photons (for the maximum beam field) corresponding to a beam lifetime of 30 minutes:
\begin{equation}
u = \eta E_0 / E_\mathrm{c} \approx 8.5; \;\;\;\;
E_\mathrm{c} \approx 0.12\eta E_0 \sim 0.1\eta E_0.
\label{7}
\end{equation}
The accuracy of this expression is quite good for any SRC because
it depends on collider parameters ($R$, $E_0$, $\sigma_z$) as well as on the assumed lifetime only logarithmically.

The critical energy is related to the beam parameters as follows:
\begin{equation}
\frac{E_\mathrm{c}}{E_0}=\frac{3\gamma {r_e}^2 N}{\alpha \sigma_x \sigma_z},
\label{10}
 \end{equation}
where $r_e=e^2/mc^2$ is the classical radius of the electron.
Combined with Eq.~\ref{7}, this imposes a restriction on
the beam parameters,
\begin{equation}
\frac{N}{\sigma_x \sigma_z} < 0.1 \eta\frac{\alpha}{3\gamma {r_e}^2},
\label{11}
\end{equation}
where $N$ is the number of particles in the beam, $\alpha=e^2/\hbar c \approx 1/137$, and $\sigma_x$ and $\sigma_z$ are the rms horizontal and longitudinal beam sizes, respectively.
{\sl This formula is the basis for the discussion that follows}. This constraint on beam parameters should be taken into account in luminosity optimization.

It can be shown that the beam lifetime given the above conditions is determined by the emission of beamstrahlung photons with energies a factor of $\sim65$ greater than $\langle E_{\gamma}\rangle$.

\section{The beam energy spread}
\noindent
In Ref.~\cite{Telnov-prl}, the rms beam energy spread due to beamstrahlung was compared to that due to synchrotron radiation in bending magnets.
It was shown that in rings with large energy acceptance the energy spread due to beamstrahlung could be larger than due to SR;
however, the lifetime is always determined by the emission of energetic single photons.

\section{Head-on and crab-waist collisions~{\rm \cite{Telnov-prl}}}
\noindent
In the ``crab waist'' collision scheme~\cite{crab,SuperB}, the beams
collide at an angle $\theta \gg \sigma_x/\sigma_z$,
in contrast with
the usual head-on collisions, where $\theta \ll \sigma_x/\sigma_z$.
The crab-waist scheme allows for
 higher luminosity when it is restricted only by the tune shift, characterized by the beam-beam
strength parameter. One should work at a beam-beam strength parameter smaller than a certain threshold value,
$\approx 0.15$ for high-energy SRCs~\cite{Zim}.

In head-on collisions, the vertical beam-beam strength parameter
(hereinafter, the ``beam-beam parameter'')
\begin{equation}
\xi_y =\frac{Nr_e\beta_y}{2\pi\gamma\sigma_x\sigma_y}\approx
\frac{Nr_e\sigma_z}{2\pi\gamma\sigma_x\sigma_y} \,\,\,\mbox{for}\,\,\, \beta_y \approx \sigma_z.
\label{xiy}
\end{equation}
In the crab-waist scheme~\cite{crab},
\begin{equation}
\xi_y =\frac{Nr_e\beta_y^2}{\pi\gamma\sigma_x\sigma_y\sigma_z}\,\,\,\mbox{for}\,\,\, \beta_y \approx \sigma_x/\theta.
\label{xiyc}
\end{equation}
The luminosity in head-on collisions
\begin{equation}
\lum \approx \frac{N^2f}{4\pi\sigma_x\sigma_y} \approx \frac{N f\gamma\xi_y }{2r_e\sigma_z};
\label{lhead}
\end{equation}
in crab-waist collisions,
\begin{equation}
\lum \approx \frac{N^2f}{2\pi\sigma_y\sigma_z \theta} \approx \frac{N^2 \beta_y f}{2\pi\sigma_x\sigma_y\sigma_z}  \approx \frac{N f \gamma\xi_y }{2r_e\beta_y}.
\label{lcrab}
\end{equation}

In the crab-waist scheme, one can make $\beta_y \ll \sigma_z$, which
enhances the luminosity by a factor of $\sigma_z/\beta_y$ compared to
head-on collisions. For example, at the proposed Italian SuperB
factory~\cite{SuperB} this enhancement could be a factor of 20--30.

\section{The beam energies where beamstrahlung is important~{\rm \cite{Telnov-prl}}}
\noindent
Using Eqs.~\ref{xiy} and \ref{xiyc} and the
restriction in Eq.~\ref{11}, we find the minimum beam energy
at which beamstrahlung becomes important. For
head-on collisions,
\begin{equation}
\gamma_\mathrm{min}=\left(\frac{0.1 \eta \alpha
  \sigma_z^2}{6\pi r_e\xi_y\sigma_y} \right)^{1/2} \propto
\frac{\sigma_z^{3/4}}{\xi_y^{1/2}\varepsilon_y^{1/4}};
\end{equation}
for crab-waist collisions,
\begin{equation}
\gamma_\mathrm{min}=\left(\frac{0.1 \eta \alpha
  \beta_y^2}{3\pi r_e\xi_y\sigma_y} \right)^{1/2} \propto
\frac{\beta_y^{3/4}}{\xi_y^{1/2}\varepsilon_y^{1/4}}.
\end{equation}

In the crab-waist scheme, beamstrahlung
becomes important at much lower energies because $\beta_y \ll
\sigma_z$. This can be understood from Eq.~\ref{xiyc}: smaller $\beta_y$
corresponds to denser beams, leading to a higher beamstrahlung rate.

For typical beam parameters presented in Table~\ref{Table1}, beamstrahlung
becomes important at energies $E_0 \gtrsim 70
\gev$ for \epem storage rings with head-on collisions; when the
crab-waist scheme is employed, this changes to the more strict $E_0
\gtrsim 20 \gev$. All newly proposed projects listed in
Table~\ref{Table1} are affected as they are designed for $E_0 \geq 120
\gev$.

\section{Luminosities for the head-on and crab-waist
schemes in a beamstrahlung-dominated regime~{\rm \cite{Telnov-prl}}}
\noindent
Now, let us find the luminosity \lum when it is restricted both by the
tune shift (beam-beam strength parameter) and beamstrahlung.
For head-on collisions,
\begin{equation}
\lum \approx \frac{(Nf)N}{4\pi\sigma_x\sigma_y}, \;
\xi_y \approx \frac{Nr_e\sigma_z}{2\pi\gamma\sigma_x\sigma_y}, \;
\frac{N}{\sigma_x \sigma_z} \equiv k \approx 0.1\eta\frac{\alpha}{3\gamma {r_e}^2}
\label{lhead-2}
\end{equation}
and $\sigma_y\approx \sqrt{\varepsilon_y \sigma_z}$.
This can be rewritten as
\begin{equation}
\lum \approx \frac{(Nf)k \sigma_z}{4\pi\sigma_y},
\;\;\;\;\; \xi_y \approx \frac{kr_e\sigma_z^2}{2\pi\gamma\sigma_y},\;\;\;\;\;\sigma_y\approx \sqrt{\varepsilon_y \sigma_z}.
\label{lhead-3}
\end{equation}

Thus, in the beamstrahlung-dominated regime the luminosity is
proportional to the bunch length, and its maximum value is determined by
the beam-beam strength parameter. Together, these equations give
\begin{equation}
\lum \approx \frac{Nf}{4\pi}\left(\frac{0.1
  \eta \alpha}{3}\right)^{2/3}\left(\frac{2\pi \xi_y}{\gamma r_e^5
  \varepsilon_y} \right)^{1/3},
\label{lhead-4}
\end{equation} \vspace{-4mm}

\begin{equation}
\sigma_{z,\mathrm{opt}}=\varepsilon_y^{1/3} \left(\frac{6\pi\gamma^2 r_e \xi_y}{0.1\eta\alpha} \right)^{2/3}.
\label{sz}
\end{equation}
Similarly, for the crab-waist collision scheme,
\begin{equation}
\lum \approx \frac{(Nf)N \beta_y
}{2\pi\sigma_x\sigma_y\sigma_z}, \;
 \xi_y \approx \frac{Nr_e\beta_y^2}{\pi\gamma\sigma_x\sigma_y\sigma_z},
\; \frac{N}{\sigma_x \sigma_z}\equiv k  \approx
0.1\eta\frac{\alpha}{3\gamma {r_e}^2}
\label{lcrab-2}
\end{equation}
and $\sigma_y\approx \sqrt{\varepsilon_y \beta_y}$. Substituting, we obtain
\begin{equation}
\lum \approx \frac{(Nf)k \beta_y}{2\pi\sigma_y}, \;\;\;\; \frac{kr_e\beta_y^2}{\pi\gamma\sigma_y} \approx \xi_y,\;\;\;\; \sigma_y\approx \sqrt{\varepsilon_y \beta_y}.
\label{lcrab-3}
\end{equation}
The corresponding solutions are
\begin{equation}
\lum \approx \frac{Nf}{4\pi}\left(\frac{0.2
  \eta \alpha}{3}\right)^{2/3}\left(\frac{2\pi \xi_y}{\gamma r_e^5
  \varepsilon_y} \right)^{1/3},
\end{equation} \vspace{-4mm}

\begin{equation}\beta_{y,\mathrm{opt}}=\varepsilon_y^{1/3}
\left(\frac{3\pi\gamma^2 r_e \xi_y}{0.1\eta\alpha} \right)^{2/3}.
\end{equation}

We have obtained a very important result: {\sl in the beamstrahlung-dominated
regime, the luminosities attainable in crab-waist and head-on collisions are
practically the same.} In fact, the gain from using the crab-waist scheme
is only a factor of $2^{2/3} \sim 1$, contrary
to the low-energy case, where the gain may be greater than one order of
magnitude. For this reason, from this point on we will consider only the case of
head-on $(\theta \ll \sigma_x/\sigma_z)$ collisions.

From the above considerations, one can find the ratio of the luminosities
with and without taking beamstrahlung into account: it is
equal to $\sigma_z/\sigma_{z,\mathrm{opt}}$ for head-on collisions and
$\beta_y/\beta_{y,\mathrm{opt}}$ for crab-waist collisions and scales
as $1/E_0^{4/3}$ for $\gamma > \gamma_\mathrm{min}$.

In practical units,
\begin{equation}
\frac{\sigma_{z,\mathrm{opt}}}{\mm} \approx
\frac{2\xi_y^{2/3}}{\eta^{2/3}} \left
(\frac{\varepsilon_y}{\nm}\right)^{1/3}
\left(\frac{E_0}{100\gev}\right)^{4/3}.
\label{sz-pract}
 \end{equation}
For example, for $\xi_y=0.15$, $\eta=0.01$, $E_0=100 \gev$ and the
vertical emittances from Table~\ref{Table1} ($\varepsilon_y=0.01$ to 0.15\nm),
we get $\sigma_{z,\mathrm{opt}}=2.5$ to 6.4\mm.

According to Eq.~\ref{lhead-4}, the maximum luminosity at high-energy
SRCs with beamstrahlung taken into account
\begin{equation}
\lum \approx
h\frac{N^2f}{4\pi\sigma_x\sigma_y}=h\frac{Nf}{4\pi}\left(\frac{0.1
  \eta \alpha}{3}\right)^{2/3}\left(\frac{2\pi \xi_y}{\gamma r_e^5
  \varepsilon_y} \right)^{1/3},
\label{19}
\end{equation}
where $h$ is the hourglass loss factor,  $f=n_\mathrm{b} c/2\pi R$ is the collision rate,
$R$ the average ring radius, and $n_\mathrm{b}$  the number of
bunches in the beam.

The energy loss by one electron in a circular orbit
$\delta E= 4\pi e^2 \gamma^4/3R_b$. Then, the power
radiated by the two beams in the ring
\begin{equation}
P=2 \delta E \frac{cNn_\mathrm{b}}{2\pi R}=\frac{4e^2\gamma^4cNn_\mathrm{b}}{3RR_\mathrm{b}}.
\label{22}
\end{equation}
Substituting $Nn_\mathrm{b}$ from Eq.~\ref{22} to Eq.~\ref{19}, we obtain
\begin{equation}
\lum \approx h\frac{(0.1\eta\alpha)^{2/3}P R}{32\pi^2\gamma^{13/3} r_e^3}
\left(\frac{R_\mathrm{b}}{R}\right)
\left(\frac{6\pi\xi_yr_e}{ \varepsilon_y} \right)^{1/3},
\label{23}
 \end{equation}
or, in practical units, \vspace{-2mm}
\begin{equation}
\frac{\lum}{10^{34}\cms} \approx \frac{100h\eta^{2/3} \xi_y^{1/3}}{(E_0/100\gev)^{13/3}
(\varepsilon_y/\nm)^{1/3} }
\left(\frac{P}{100\, \mbox{MW}}\right)
\left(\frac{2\pi R}{100\km}\right) \frac{R_\mathrm{b}}{R}. \vspace{-1mm}
\label{29}
\end{equation}

Once the vertical emittance is given as an input parameter, we find
the luminosity and the optimum bunch length by applying
Eq.~\ref{sz-pract}.  Beamstrahlung and the beam-beam strength
parameter determine only the combination $N/\sigma_x$; additional
technical arguments are needed to find $N$ and $\sigma_x$ separately.
When they are fixed, the optimal number of bunches $n_\mathrm{b}$ is
found from the total SR power, Eq.~\ref{22}.

In Table~\ref{Table2}, we present the realistic luminosities
and beam parameters for the rings listed in Table~\ref{Table1}
after both beamstrahlung and the beam-beam parameter are taken into account.
The following assumptions are made:
SR power $P=100$ MW,
$R_\mathrm{b}/R=0.7$,
the hourglass factor $h=0.8$, the beam-beam parameter $\xi_y=0.15$,
the energy acceptance of rings $\eta=0.01$;
the values of all other parameters ($\varepsilon_y$, $\varepsilon_x$ and
$\beta_x$) are taken from Table~\ref{Table1}.

\begin{table*}[tbh]
\vspace{0mm}
{\renewcommand{\arraystretch}{0.88}
\small
\begin{tabular}{l c c c c c c c c c c c}
& LEP & LEP3 & DLEP & STR1 & STR2 & STR3 & STR4 & STR5 & STR6 \\[-1mm]
&&&&&&cr-w&cr-w&cr-w&cr-w \\  \hline \\[-2mm]
$2E_0$, \gev & 209 & 240 & 240 & 240 & 240 & 240 & 400 & 400 & 500 \\
Circumference,\km & 27& 27 & 53 & 40 & 60 & 40 & 40 & 60 & 80 \\
Bunches/beam &$\sim 2$ &$\sim 7$& 70 & 24 & 53 & 240 & 36 & 45 & 31 \\
$N,\;10^{11}$  & 33 & 5.9 & 2.35 & 3.9 & 4. & 0.4 & 0.34 & 0.6& 0.65  \\
$\sigma_{z}$, \mm  & 8.1 &8.1 & 5.7 & 6.9 & 6.9 & 3.4 & 6.7 & 7.8 &9.6  \\
$\sigma_y$, \mum  & 1.4 & 1.1 & 0.53 & 0.78 & 0.78 & 0.19 & 0.27 & 0.36 &0.35  \\
\lum, $10^{34}\cms$ & 0.47 & 0.31 & 0.89 & 0.55 & 0.83 & 1.1 & 0.12 & 0.16 & 0.087
\end{tabular}
\vspace{0mm}
}
\caption{Realistically achievable luminosities and other beam parameters for the
projects listed in Table~1 at synchrotron-radiation power $P=100$ MW.
Only the parameters that differ from those in Table~1 are shown.}
\label{Table2}
\end{table*}

Comparing Tables~\ref{Table1} and \ref{Table2}, one can see that
at $2E_0=240 \gev$, which is the preferred energy for the study of
an $m=126 \gev$ Higgs boson, taking beamstrahlung into account
lowers the luminosities achievable at storage-ring colliders with
crab-waist collisions by a factor of 15. Nevertheless, these luminosities are comparable
to those at the ILC, $\lum_{\mathrm{ILC}} \approx
(0.55$--$0.7) \times 10^{34} \cms$ at $2E_0=240 \gev$~\cite{ILCinterim}.
However, at $2E_0=500 \gev$ the ILC can achieve
$\lum_{\mathrm{ILC}} \approx (1.5$--$2) \times 10^{34} \cms$,
which is a factor 15--25 greater than the luminosities achievable
at storage rings.

\section{Ways to increase luminosity of high-energy storage rings}

\subsection{Linear-ring collider with energy recuperation}
\begin{figure}[ht]
     \begin{center}
     \vspace*{-0.2cm}
     \includegraphics[width=14cm] {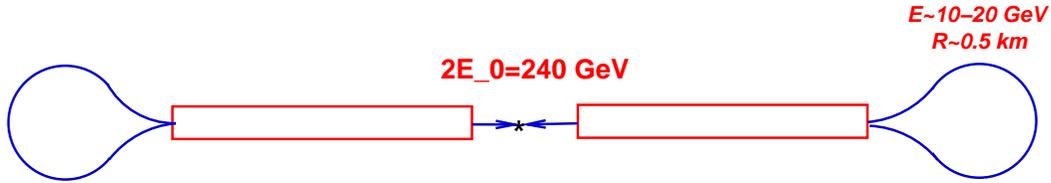}
    \vspace*{-0.3cm}
     \end{center}
     \caption{Linear-ring collider with energy recuperation}
   \vspace*{0.5cm}
   \label{rec}
   \end{figure}

\noindent
At linear colliders, beams are used only once; one should accelerate new
beams from zero to the maximum energy for each beam collisions; therefore,
LCs are very energy-ineffective machines. Storage rings are much better
in this respect; however, at high beam energies they also need
huge electrical power to compensate for SR energy losses. At first sight,
these problems can be partially overcome in the collider scheme shown
in Fig.~\ref{rec}. In this scheme, after the collision at the interaction
point (IP), the electron (positron) bunch  is decelerated in the second
half of the linac down to the energy of about 15 \gev, makes one turn in the
ring, then is accelerated again---and the process is repeated over and over again
(for minutes) as in conventional storage rings but with much smaller SR losses
(proportional to $E^4/R$ per turn).

If $\eta$ is the energy acceptance of the ring, the maximum energy of
beamstrahlung photons should be $\eta E$ (not $\eta E_0$). This reduces
luminosity $\lum$ by a factor of $(E/E_0)^{2/3} \sim 0.25$. However, thanks to much lower
SR losses, one can increase $Nf$ by a very large factor, and thus
increase the luminosity by 1-2 orders of magnitude.

     Unfortunately, there are many stoppers which make this scheme impractical:
\begin{itemize}
\item The required refrigeration power for the linac working in the continuous mode is about 150--200 MW (assuming an acceleration gradient
of $\sim 15$ MeV/m, $Q=2\times 10^{10}$)~\cite{ERL}.
\item Parasitic collisions of beams inside the linac. One can separate the beams (the pretzel scheme), but beam
attraction would lead to beam instabilities.
\item The transverse wakefield problem for beams shifted from the axis.
\item The energy difference between the head and tail becomes unacceptable after deceleration (beam loading helps during acceleration but is detrimental during deceleration).
\end{itemize}
         Thus, this idea looks interesting but technically impossible. LC schemes with recuperation were considered in 1970s~\cite{Keil}
         and were also rejected then.

\subsection{Charge-compensated (four-beam) collisions}
\noindent
The idea to collide four beams (\epem with \epem) is more than 40 years old.
Beams are neutral at the IP, so there are no collision effects, which sounds nice.
A four-beam \epem\ collider, DCI, with the energy $2E \sim 2$ GeV, was build
in 1970s in Orsay~\cite{DCI}. There were hopes the four-beam collision scheme would increase
the luminosity by a factor of 100---but the result was confusing: the maximum luminosity was
approximately the same as in two-beam collisions. This is due to
the instability of neutral \epem\ beams~\cite{Derbenev}: a small displacement of charges in one
beam leads to charge separation in the opposing beam and subsequent
development of beam instabilities as in two-beam collisions. The
beam-beam parameter $\xi_y$ attainable at DCI was approximately the same as
without neutralization.

   A similar four-beam approach was considered in 1980s for linear colliders
in order to suppress beamstrahlung~\cite{Solyak}. However, simulation and theory have
shown that the kink instability that develops in collision limits the
luminosity to a value that is lower than in two-beam collisions because
the pinch effect is absent; beamstrahlung is suppressed only at rather low
luminosities. For these reasons, the idea of charge neutralization at
\epem colliders was dismissed. Why are we discussing it again?

The above-mentioned beam neutralization exercises have shown that this
method does not help to increase the beam-beam parameter $\xi$. However, it should
work at values of $\xi_y$ attainable in two-beam collisions. We have such
a case at \epem Higgs factories. Their luminosities are limited by
beamstrahlung. Using charge compensation, one can suppress
beamstrahlung and increase the luminosity while keeping the same $\xi_y$ as for
standard two-beam collisions.

What luminosity increase can be achieved this way? For head-on collisions,
beam neutralization can increase the luminosity only
by a factor of 2--3 at $2E_0=240$ GeV. However, in the
crab-waist scheme (which, as demonstrated above, has no benefit in beamstrahlung-dominated
ring colliders) the situation is much more attractive. Comparing
the luminosities for crab-waist storage rings in Tables~\ref{Table1} and \ref{Table2} (without and with
beamstrahlung taken into effect), one can see the possible benefits from the suppression of
beamstrahlung (in the inline table below, the subscript ``nb" stands for ``no beamstrahlung'', ``b" stands for ``with beamstrahlung''):
\begin{center}
\begin{tabular}{l | c c c c }
$2E_0, \gev$ & 240 & 400 & 400 & 500  \\[-1mm]
\hline
$\lum_{\rm nb}/\lum_{\rm b}$ & 16 & 33 & 43 & 25 \\
\end{tabular}
\end{center}
These numbers, corresponding to an ideal charge compensation, look quite
attractive.

Let us find the charge compensation ``quality'' required to increase the luminosity by a factor of $\lum_\c/\lum$,
where the subscript ``\c'' denotes ``compensated''. From Eqs.~\ref{lcrab-2} and \ref{lcrab-3}, we have
\begin{equation}
\lum \propto k^{2/3}\xi_y^{1/3},
\label{n-comp}
\end{equation}
where $k \equiv N/(\sigma_x \sigma_z) \approx 0.1\eta \alpha/(3\gamma
{r_e}^2)$. For the case of charge-compensated beams, the beamstrahlung
condition should be rewritten as follows:  $\Delta N_\c/(\sigma_{x,\c}
\sigma_{z,\c}) \equiv k$ or $N_\c/(\sigma_{x,\c} \sigma_{z,\c}) \equiv
k^{\prime}=k(N_\c/\Delta N_\c)$. The expression for luminosity will be
similar to Eq.~\ref{n-comp}, with $k$ replaced by $k^{\prime}$:
\begin{equation}
\lum_c \propto \left(k\frac{N_\c}{\Delta N_\c}\right)^{2/3}\xi_{y,\c}^{1/3}.
\label{comp}
\end{equation}
From these two equations, we get the required degree of charge compensation
\begin{equation}
\frac{\Delta N_\c}{N_\c}=\left(\frac{\lum}{\lum_\c}\right)^{3/2} \left(\frac{\xi_\c}{\xi}\right)^{1/2}.
\label{dn}
\end{equation}
For example, if $\xi_\c=\xi$ , then to increase the luminosity by
a factor of $\lum_\c/\lum=10$ one needs $\Delta N_\c/N_\c=0.03$. Gaining a factor of 30
requires $\Delta N/N=0.006$. Thus, an order-of-magnitude luminosity increase
seems realistic.
Unfortunately, this method has many problems (and even stoppers):

 \begin{figure}[tbh]
     \begin{center}
     \vspace*{0.2cm}
     \includegraphics[width=15cm] {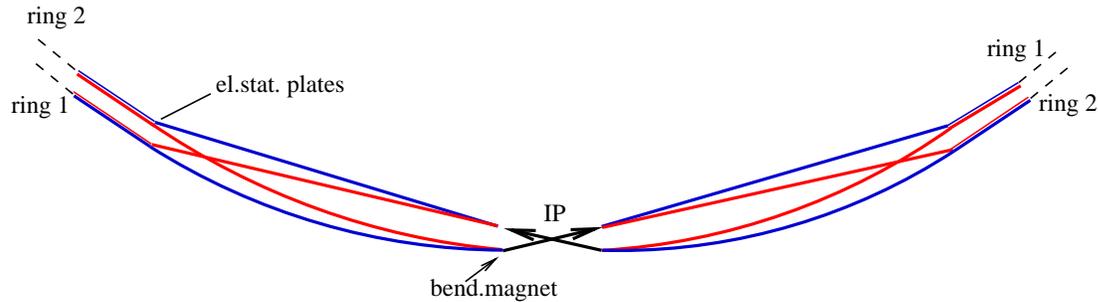}
    \vspace*{-0.3cm}
     \end{center}
     \caption{Ring collider with charge compensation. Blue: electron, red: positron beamlines.}
   \vspace*{0.0cm}
   \label{2rings}
   \end{figure}

\begin{enumerate}
\item Crab-waist collisions assume collisions at some horizontal angle, which
requires two rings with electrostatic separators near the IP and inside the rings, see Fig.~\ref{2rings}, plus one
ring for injection; three rings in total.
\item The assembly of electron and positron bunches approaching the IP at a relative angle of about 10 mrad
(determined by quads radius) into one neutral bunch using a bending magnet placed between the IP and
the final-focus quads looks very problematic due to synchrotron radiation in the magnet. This leads to
the increase of the vertical beam size at the IP (deflection in the solenoidal detector field due to the crossing angle).
This problem is very serious and appears to be a stopper for high-energy rings.
\item Everything must be very close to ideal: even a small displacement of beams at the IP (e.g.,
$0.2\sigma_y$) would lead to a very short beam lifetime.
\item Immediately after refilling, bunches have slightly different sizes and some displacement.
One should avoid collisions of such bunches at the IP until full damping has been achieved.
\item The additional cost.
\end{enumerate}

\section{Conclusion}
\noindent
We have shown that the beamstrahlung
phenomenon must be properly taken into account in the design and
optimization of high-energy storage rings colliders
(SRCs). Beamstrahlung determines the beam lifetime through the emission
of single photons in the tail of the beamstrahlung spectra, thus severely limiting the
luminosity. We have demonstrated that beamstrahlung suppresses the
luminosities of high-energy \epem storage rings as $1/E_0^{4/3}$ at
beam energies $E_0 \gtrsim 70 \gev$ for head-on collisions and $E_0
\gtrsim 20 \gev$ for crab-waist collisions. Very importantly,
beamstrahlung makes the luminosities attainable in head-on and
crab-waist collisions approximately equal above these threshold
energies.  At $2E_0 = 240$--$500 \gev$, beamstrahlung lowers the
luminosity of crab-waist rings by a factor of 15--40. Some increase in
SRC luminosities can be achieved at rings with larger radius, larger
energy acceptance, and smaller beam vertical emittance.

We also conclude that the luminosities attainable at \epem storage
rings (with one interaction point) and linear colliders are comparable
at $2E_0 = 240 \gev$.  However, at $2E_0 = 500 \gev$ storage-ring
luminosities are by a factor of 15--25 smaller (this factor is smaller for rings with larger radius).
Therefore, linear colliders remain the most promising instrument for
studying the physics at energies $2E_0 \gtrsim 250 \gev$.

We have also briefly considered two possible methods of increasing the luminosity of high-energy \epem storage rings:
the linac-ring storage ring scheme with the energy recuperation and the charge-compensated four-beam collider.
The first scheme looks completely unrealistic; the second one is also very difficult to implement
and suffers from a fundamental problem connected with radiation in beam-combining magnets.

\section*{Acknowledgement}
\noindent
I would like to thank A.~Blondel, F.~Zimmermann and K.~Oide for the revival of interest to high-energy \epem storage rings. Many thanks to M.~Koratzinos, E.~Levichev,  A.~Skrinsky, K.~Yokoya, and M.~Zanetti for their interest in the beamstrahlung problem, and to A.~Miagkov for the invitation to this workshop.\\

\noindent
This work was supported by Russian Ministry of Education and Science.

\end{document}